\documentclass[runningheads,orivec]{llncs}
\usepackage[T1]{fontenc}
\usepackage{amsmath}
\usepackage{amsfonts}
\usepackage{graphicx}
\usepackage{adjustbox}
\usepackage{textcomp}
\usepackage{listings}
\usepackage{booktabs}
\usepackage{xcolor}
\usepackage{hyperref}
\usepackage[disable]{todonotes}
\usepackage{tikz}
\usetikzlibrary{decorations.pathreplacing, positioning}
\usepackage{soul}
\usepackage{arydshln}
\definecolor{keywords}{RGB}{0,155,155}
\definecolor{comments}{RGB}{155,0,0}
\definecolor{string}{RGB}{70,145,90}
\definecolor{green}{RGB}{70,145,90}
\definecolor{lblue}{RGB}{240,240,255}
\definecolor{rewritehl}{RGB}{255,228,235}

\usepackage{color-edits}
\addauthor[Michael]{mg}{cyan}

\begin{document}

\title{Compiling Rewrite Rules to Finite-State Transducers with the Worsening Trick}
\titlerunning{Compiling Rewrite Rules with Worsening}

\newcommand{\notex}[4][]{\todo[author=#2,color=#3,size=\scriptsize,fancyline,caption={},#1]{#4}} 
\newcommand{\mans}[2][]{\notex[#1]{mans}{blue!40}{#2}}
\newcommand{\Mans}[2][]{\mans[inline,#1]{#2}\noindentaftertodo}

\newcommand{\michael}[2][]{\notex[#1]{michael}{orange!40}{#2}}
\newcommand{\Michael}[2][]{\michael[inline,#1]{#2}\noindentaftertodo}

\author{Mans Hulden\inst{1} \and
Michael Ginn\inst{2}}

\institute{New College of Florida, Sarasota FL 34243, USA\\
\email{mhulden@ncf.edu}\\
\and
University of Colorado, Boulder CO 80309, USA\\
\email{michael.ginn@colorado.edu}}

\authorrunning{M. Hulden and M. Ginn}
\maketitle

\begin{abstract}
  Finite-state transducers (FSTs) are essential for modeling string rewriting in computational linguistics and natural language processing (NLP), particularly for phonological and morphological rewrite rules. Compiling general rewrite rules of the form $A \to B / L \, \_ \, R$, where $A$, $B$, $L$, and $R$ are arbitrary regular languages, is complex due to overlapping matches and context constraints. Traditional methods, such as those by Kaplan and Kay or Karttunen, rely on intricate transducer compositions with auxiliary markers. This paper presents a compact compilation scheme based on the ``worsening trick'' \cite{gerdemann-van-noord-2000-approximation,gerdemann2012practical}: generate all legal rewrite candidates, then filter candidates that are worse than another candidate for the same input. Implemented as the built-in rewrite compiler in PyFoma \cite{hulden-etal-2024-pyfoma}, the construction supports multiple contexts, arbitrary transductions, markup, directed rewriting, weights, and parallel rewriting. The resulting formulas are short and uniform, and where semantics coincide, they reproduce the same rule transducers as earlier approaches while remaining easier to extend. The implementation has been validated against {\it foma} \cite{hulden2009foma} on both a substantial collection of rewrite grammars and an automated regression suite covering the major rewrite modalities, with the resulting transducers matching exactly apart from state numbering.

\end{abstract}

\section{Introduction}

String rewriting is fundamental to computational linguistics, particularly in phonology and morphology, where rewrite rules transform strings based on contextual constraints \cite{chomsky1968sound,johnson1972formal}. A general rewrite rule, written as $A \to B / L \, \_ \, R$, replaces strings from a regular language $A$ with $B$ in the context $L \, \_ \, R$, where $L$ and $R$ are also regular languages, while copying unmatched input substrings to the output. Compiling such rules into finite-state transducers (FSTs) enables efficient processing but is challenging due to overlapping matches and multiple contexts, especially when $A$, $B$, $L$, and $R$ are arbitrary regular languages \cite{kaplan1994regular}. The difficulty lies in the fact that material targeted for rewriting can simultaneously serve as context for another site of rule application \cite{karttunen1995replace}.

We present a compilation method based on the ``worsening trick'' \cite{gerdemann-van-noord-2000-approximation,gerdemann2012practical}. We first generate all candidates that mark possible rewrite sites, then restrict those candidates to the legal contexts, and finally remove every candidate that can be made strictly worse by deleting or shifting some rewrite marking. The same three-stage pattern handles obligatory rewriting, leftmost/longest/shortest preferences, arbitrary transductions, markup, parallel rules, and weights.\footnote{A JavaScript port of PyFoma is available for interactive experimentation in the browser at \url{https://fomafst.github.io/pyfomajs/}. It implements the rewrite compiler described here and can visualize the resulting machines. PyFoma is available at \url{https://github.com/mhulden/pyfoma} and \url{https://pypi.org/project/pyfoma/}.}

Our contribution is a formulation in which the compilation itself can be stated with substantially shorter and more uniform formulas than in marker-heavy classical presentations, while still matching the established semantics. The paper is organized as follows: Section \ref{sec:rewrite_rules} states the rule semantics assumed throughout, Section \ref{sec:worsening} reviews worsening, Section \ref{sec:compiling_rewrite_rules} gives the construction, and Section \ref{sec:conclusion} concludes. A readable reference implementation in PyFoma, closely following the constructions in the paper, is given in the appendix (\ref{ref:pyfoma_implementation}).

\section{Related Work}

Foundational work by Kaplan and Kay \cite{kaplan1994regular} introduced a method for compiling rewrite rules by inserting auxiliary markers, constraining their distribution, performing the relevant replacements, and finally removing the markers again. Subsequent approaches by Karttunen \cite{karttunen1995replace,karttunen1996directed}, Kempe and Karttunen \cite{kempe1996parallel}, and Mohri and Sproat \cite{mohri1996efficient} refined this general marker-based strategy, incorporating directed rewriting, parallel rewriting, and weights. The Pynini treatment described by Gorman and Sproat \cite{gorman2022finite} follows this same general line, presenting rewrite compilation as a composition of several auxiliary transducers with marker symbols and boundary-handling machinery.

A different attempt to make rewrite compilation conceptually clearer is {\it foma}'s multitape treatment \cite{hulden2009finite}, where an overgenerating multitape automaton is first constructed and then filtered so that only licensed rewrites remain before projection to an ordinary two-tape transducer. This idea also underlies other practical systems such as Kleene \cite{beesley-2012-kleene}. Efforts by Vaillette \cite{vaillette2004logical} and Hulden \cite{hulden2009regular} use predicate logic to simplify correctness arguments, but the resulting rewrite constructions still require substantial auxiliary machinery.


The worsening trick itself has already been used for regular sets and transductions \cite{gerdemann-van-noord-2000-approximation,gerdemann2012practical}. Gerdemann \cite{gerdemann-2009-mix} sketches a declarative rewrite implementation based on worsening using star-marked candidates and a flattening step. In contrast, our contribution is a complete, practical compilation scheme for the classical rule format $A \to B / L \, \_ \, R$ that operates entirely within ordinary regular languages and relations. The construction factors cleanly into three transparent steps: (i) generate all candidate rewrites by bracketing possible transduction sites in $T_{\rm base}$, (ii) filter out candidates containing rewrite sites in illicit contexts using the dotted-position restriction technique of Yli-Jyrä and Koskenniemi, and (iii) retain only maximal candidates under a worsening relation that encodes obligatory application and directionality preferences. Where the semantics coincide, the resulting transducers are identical (modulo state renumbering) to those produced by the established {\it foma} compiler, yet the formulas are substantially shorter and more uniform than both classical marker-based constructions and multitape approaches, and significantly easier to extend to weights, parallel rules, markup, epsilon rules, and output-side context checking.

\section{Implementation and Validation}

Our construction is implemented in PyFoma, and we validated it against the C-based {\it foma} rule compiler both on a collection of 31 grammars comprising 427 rules\footnote{A subset of these grammars is publicly available on the PyFoma JavaScript port site.} and on an automated regression suite covering the major rewrite modalities shared by the two systems, including ordinary contextual rewriting, directed rewriting, insertion, markup, and parallel rewriting. At the time of writing, the regression suite contains 2,217 individual cases. In all tested cases the resulting transducers were structurally identical, differing only in state numbering. This is consistent with the fact that both the worsening construction and {\it foma}'s multitape-based compiler build the same basic pattern of identity transitions outside rewrite centers and canonical cross-product transductions inside rewrite spans, so under standard Hopcroft-style minimization \cite{hopcroft1971} over input:output label pairs they yield the same result in the shared cases tested here.

\section{Notation}

We work over an alphabet $\Sigma$ of ordinary symbols and use auxiliary symbols $\{\#,<,>\}$, disjoint from $\Sigma$, to mark word boundaries and candidate rewrite sites. Let $\Gamma = \Sigma \cup \{\#, <, >\}$. Intermediate languages are subsets of $\Gamma^*$, and intermediate transducers are relations in $\Gamma^* \times \Gamma^*$. When convenient, we identify a language $L \subseteq \Gamma^*$ with its identity relation $\{(x,x) \mid x \in L\}$. The final compiled rule transducers, however, relate strings over $\Sigma$ only.

Composition is denoted by $\circ$, with the convention
\[
T_1 \circ T_2 = \{(x, z) \mid \exists y \, ((x, y) \in T_1 \land (y, z) \in T_2)\}.
\]
For regular languages $A,B$, we express the cross product as
\[
A\!:\!B = \{(x,y) \mid x \in A, y \in B\},
\]
that is, the regular relation pairing every string in $A$ with every string in $B$. The complement of a language $L \subseteq \Gamma^*$ is $\neg L = \Gamma^* - L$. For a relation $T$, $\text{range}(T) = \{y \mid \exists x \, (x,y)\in T\}$ and $\text{dom}(T) = \{x \mid \exists y \, (x,y)\in T\}$.

PyFoma examples use the regular-expression syntax shown in Table \ref{tab:syntax}. In the transducer diagrams (Figure~\ref{fig:tbase} and later), a {\bf .} on a transition denotes the identity transition on any symbol not explicitly included in that transducer's alphabet, and multiple identical transitions are merged into one with comma-separated labels.

\begin{table}[ht]
    \centering
    \begin{tabular}{l  c c}
        \toprule
       {\bf Operator} & {\bf PyFoma}  & {\bf Regular expression}\\
       \midrule
        union & {\tt $\mid$} & $\cup$ \\
        intersection & ${\tt \&}$ & $\cap$ \\
        cross-product & $:$ & $:$ \\
        Kleene star & $*$ & $*$ \\ 
        Kleene plus & $+$ & $+$ \\
        wildcard & ${\tt .}$ & $\Sigma$ \\ 
        epsilon & {\tt \textquotesingle \textquotesingle} & $\epsilon$ \\
        subtraction & ${\tt -}$ & $-$ \\
        composition & ${\tt @}$ & $\circ$ \\
        character class & ${\tt [a-z]}$ & N/A \\
        negated character class & \texttt{[\char`\^a-z]} & N/A \\
        domain & {\tt \$\^{}input(T)} & dom(T) \\
        range & {\tt \$\^{}output(T)} & range(T) \\
        rewrite rule & {\tt \$\^{}rewrite(A:B / L \_ R)} & $A \to B / L \, \_ \, R$ \\

        \bottomrule
        \addlinespace[1.0ex]
    \end{tabular}
    \caption{PyFoma and standard regular expression syntax for various finite-state operators.}
    \label{tab:syntax}
\end{table}

When $\Sigma$ occurs inside a transducer expression, we use it as shorthand for the identity relation on symbols from $\Sigma$.

\section{Rewrite Rules}
\label{sec:rewrite_rules}

Rewrite rules originated in phonological theory, notably in Chomsky and Halle's \textit{The Sound Pattern of English} \cite{chomsky1968sound}, where rules like $b \to p / \, \_ \, \#~ $ devoice a $b$ word-finally. Johnson \cite{johnson1972formal} showed that such rules, when contexts are regular languages, can be modeled as FSTs, if they are assumed not to reapply to their own outputs, unlike context-sensitive grammar rules.

A general rewrite rule is written as:

$$
A \to B / L_1 \, \_ \, R_1 , \dots , L_n \, \_ \, R_n,
$$

\noindent where $A$, $B$, $L_i$, and $R_i$ are regular languages, and {\#} denotes a word edge in the contexts. For example, the rule $b \to p / \, \_ \, \#, \# \_$ would turn a $b$ to a $p$ either word-initially or word-finally.

We assume the standard simultaneous interpretation of rewrite rules. A rule application chooses a possibly empty set of non-overlapping substrings of the input, each belonging to $A$ and satisfying at least one listed context $L_i \, \_ \, R_i$ on the input side, replaces each chosen substring once by its image under $A\!:\!B$, and copies all remaining symbols unchanged. Thus the rule defines a relation between whole input strings and whole output strings; it is not interpreted as an iterative procedure that repeatedly reapplies the rule to its own output. Later, in Section \ref{sec:output_side}, we discuss the alternative variant where contexts are checked on the output side. If several distinct sets of legal occurrences exist, the rule is nondeterministic unless further preferences such as leftmost or longest are imposed.

For example, the rule $a^+ \to x / a \, \_ \, a$ replaces sequences of one or more $a$ with $x$ when flanked by $a$ on both sides. An input string $aaa$ rewrites to $axa$. For $aaaa$, the legal simultaneous applications yield $\{axxa, axa\}$: either the two middle $a$'s are rewritten separately, or the middle $aa$ is rewritten as one span. Outputs such as $axaa$ or $aaxa$ omit a licensed rewrite site and are therefore excluded under obligatory simultaneous semantics. If we additionally require longest-match behavior, then only $axa$ is produced from $aaaa$.

\section{Worsening}
\label{sec:worsening}

\subsection{The worsening trick}

\cite{gerdemann-van-noord-2000-approximation,gerdemann2012practical} use a strategy of imposing a preference relation on strings in a regular language to filter out candidate strings from the same set deemed to be `suboptimal'. To illustrate the general technique, consider a language $L_m$ containing strings with optional morpheme boundaries (a $+$-symbol), e.g., $\{$de+construct+ion+s, deconstruct+ion+s, deconstructions, $\ldots$ $\}$. To select those strings in $L_m$ with the maximal number of $+$ symbols, define a worsening transducer $W$ that removes at least one $+$ (but may remove more):

\[
W = (\Sigma^*~ (+\!:\!\epsilon)~ \Sigma^*)^+,
\]

where $\epsilon$ is the empty string. The language of strings with maximal $+$ symbols is:

\[
\max_{+}(L_m) = L_m \cap \neg \text{range}(L_m \circ W),
\]

\noindent where $\text{range}(T)$ denotes the output language of transducer $T$. This filters out strings from $L_m$ that can be derived by removing $+$-symbols, retaining only those with the most $+$-symbols, e.g., $\text{de+construct+ion+s}$.

For transducers, suppose a transducer $T$ generates candidate transductions, and we want to filter its output. Define a worsening transducer $W$ that encodes a preference relation among competing outputs, and compute:

\[
T \circ \neg \text{range}(T \circ W),
\]

\noindent or, symmetrically

\[
\neg \text{range}(\text{dom}(T) \circ W) \circ T,
\]

\noindent if we want to filter out strings from the input language of a transducer $T$. The key idea is that the worsener does not describe the bad candidates directly; instead, it maps better candidates to worse competing candidates, which can then be removed algebraically in one step. Here the language $\neg \text{range}(\cdots)$ is understood as its identity relation when composed with a transducer.

\begin{figure}
\centering
\begin{tikzpicture}[scale=0.55, every node/.style={scale=0.95}]
  \node[fill=red!20, font=\bfseries] (top) at (0, 5.2) {\texttt{<a><a><a>}};

  \node (l2a) at (-5.2, 4.0) {\texttt{<a><a>a}};
  \node (l2b) at (-2.6, 4.0) {\texttt{<a>a<a>}};
  \node (l2c) at (0, 4.0) {\texttt{a<a><a>}};
  \node[fill=red!20, font=\bfseries] (l2d) at (2.6, 4.0) {\texttt{<a><aa>}};
  \node[fill=red!20, font=\bfseries] (l2e) at (5.2, 4.0) {\texttt{<aa><a>}};

  \node (l1a) at (-6.8, 2.6) {\texttt{<a>aa}};
  \node (l1b) at (-4.1, 2.6) {\texttt{a<a>a}};
  \node (l1c) at (-1.3, 2.6) {\texttt{aa<a>}};
  \node (l1d) at (1.3, 2.6) {\texttt{<aa>a}};
  \node (l1e) at (4.1, 2.6) {\texttt{a<aa>}};
  \node[fill=red!20, font=\bfseries] (l1f) at (6.8, 2.6) {\texttt{<aaa>}};

  \node (bottom) at (0, 1.1) {\texttt{aaa}};

  \draw (top) -- (l2a);
  \draw (top) -- (l2b);
  \draw (top) -- (l2c);
  \draw (l2a) -- (l1a); 
  \draw (l2a) -- (l1b); 
  \draw (l2b) -- (l1a); 
  \draw (l2b) -- (l1c); 
  \draw (l2c) -- (l1b); 
  \draw (l2c) -- (l1c); 
  \draw (l2d) -- (l1a); 
  \draw (l2d) -- (l1e); 
  \draw (l2e) -- (l1c); 
  \draw (l2e) -- (l1d); 
  \draw (l1a) -- (bottom);
  \draw (l1b) -- (bottom);
  \draw (l1c) -- (bottom);
  \draw (l1d) -- (bottom);
  \draw (l1e) -- (bottom);
  \draw (l1f) -- (bottom);
\end{tikzpicture}
\caption{Hasse-style diagram of the ordering induced by the obligatory worsener for the rewrite rule $a^+ \to x$ on input $aaa$. Nodes represent candidate strings; edges show selected one-step worsenings obtained by removing a single bracket pair, rather than every comparable pair in the order. The maximal elements are \texttt{<a><a><a>}, \texttt{<a><aa>}, \texttt{<aa><a>}, and \texttt{<aaa>}; they survive filtering and produce the outputs $xxx$, $xx$, $xx$, and $x$, respectively.}
\label{fig:hasse}
\end{figure}
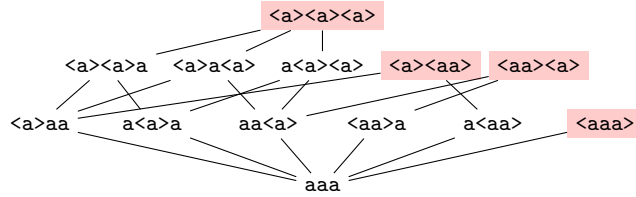

This retains only outputs that cannot be worsened, forming the core of our compilation strategy. Figure \ref{fig:hasse} illustrates the idea.

\section{Compiling Rewrite Rules---Basic Cases}
\label{sec:compiling_rewrite_rules}

We compile a rewrite rule $A \to B / L \, \_ \, R$ into an FST as follows using the notation introduced above. Here, $T$ denotes the center transduction $A\!:\!B$, a cross-product of two regular languages, though we will later make $T$ more general (see \ref{sec:arb}).

\subsection{Base Transducer}

Generate all possible rewrite candidates with:

\begin{equation}
T_{base} = \# \, (\Sigma~ \cup < \, T \, >)^* \, \#,
\end{equation}

\noindent where $\Sigma$ represents non-rewritten symbols, and $< \, T \, >$ marks a rewrite site where strings from $A$ are replaced by strings from $B$. 

\begin{figure}
\begin{center}
\begin{adjustbox}{scale=0.9}
 \begin{tikzpicture}[scale=0.2]
\tikzset{
    state/.style={circle, draw=black, fill=gray!30, inner sep=0pt, minimum size=1.2cm},
    final/.style={double, double distance=1.2pt, fill=gray!30} 
}

\draw [black] (7.3,-31.7) -- (9.3,-31.7);
\fill [black] (9.3,-31.7) -- (8.5,-31.2) -- (8.5,-32.2);

\draw [black, fill=gray!30] (12.3,-31.7) circle (3);
\draw (12.3,-31.7) node {$0$};

\draw [black, fill=gray!30] (23.8,-31.7) circle (3);
\draw (23.8,-31.7) node {$1$};

\draw [black] (23.8,-41.9) circle (3);
\draw [black, fill=gray!30] (23.8,-41.9) circle (2.4);
\draw (23.8,-41.9) node {$2$};

\draw [black, fill=gray!30] (34.4,-31.7) circle (3);
\draw (34.4,-31.7) node {$3$};

\draw [black, fill=gray!30] (45.3,-31.7) circle (3);
\draw (45.3,-31.7) node {$4$};

\draw [black, fill=gray!30] (56.4,-31.7) circle (3);
\draw (56.4,-31.7) node {$5$};

\draw [black] (15.3,-31.7) -- (20.8,-31.7);
\fill [black] (20.8,-31.7) -- (20,-31.2) -- (20,-32.2);
\draw (18.05,-31.2) node [above] {$\#\!:\!\#$};

\draw [black] (22.477,-29.02) arc (234:-54:2.25);
\draw (23.8,-24.45) node [above] {$.,\mbox{ }a\!:\!a,\mbox{ }b\!:\!b,\mbox{ }x\!:\!x$};
\fill [black] (25.12,-29.02) -- (26,-28.67) -- (25.19,-28.08);

\draw [black] (23.8,-34.7) -- (23.8,-38.9);
\fill [black] (23.8,-38.9) -- (24.3,-38.1) -- (23.3,-38.1);
\draw (23.3,-36.8) node [left] {$\#\!:\!\#$};

\draw [black] (26.8,-31.7) -- (31.4,-31.7);
\fill [black] (31.4,-31.7) -- (30.6,-31.2) -- (30.6,-32.2);
\draw (29.1,-31.2) node [above] {$<:<$};

\draw [black] (37.4,-31.7) -- (42.3,-31.7);
\fill [black] (42.3,-31.7) -- (41.5,-31.2) -- (41.5,-32.2);
\draw (39.85,-31.2) node [above] {$a\!:\!x$};

\draw [black] (48.3,-31.7) -- (53.4,-31.7);
\fill [black] (53.4,-31.7) -- (52.6,-31.2) -- (52.6,-32.2);
\draw (50.85,-31.2) node [above] {$b\!:\!\epsilon$};

\draw [black] (53.912,-33.373) arc (-59.26772:-120.73228:27.027);
\fill [black] (26.29,-33.37) -- (26.72,-34.21) -- (27.23,-33.35);
\draw (40.1,-37.67) node [below] {$>:>$};

\end{tikzpicture}
\end{adjustbox}
\end{center}
 \caption{$T_{\rm base}$ for the rule $ab \to x$. This transducer is independent of the contexts $L$ and $R$; the same machine is used for any context and the restriction to valid sites is enforced later by composition with $L_{\rm context}$. The symbol {\bf .} denotes PyFoma's catch-all transition label (identity on any symbol not explicitly listed in the current alphabet).}
    \label{fig:tbase}

\end{figure}
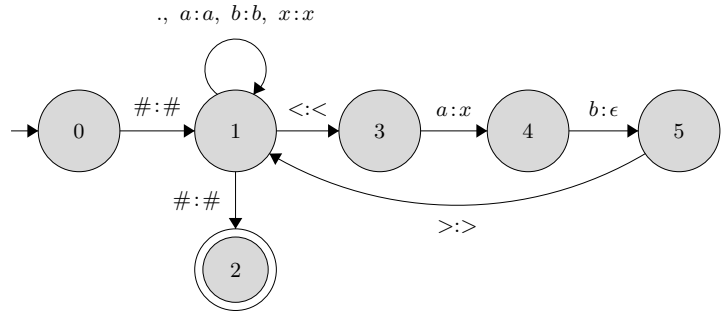

A typical path through $T_{base}$ corresponds to a candidate such as $\# \, a \, < \, a:\!x ~ b:\!\epsilon > \, a \, \#$, representing a rewrite of $ab$ to $x$, with other symbols in an identity relation (see Figure \ref{fig:tbase}). We explicitly insert \textbf{\#} symbols at the beginning and end of every string so that word-edge contexts (such as {\_ \#} or {\# \_}) can be treated uniformly as ordinary left- and right-contexts in the same restriction mechanism---no special handling for string boundaries is required.

\subsection{Context Restriction}

We then restrict bracket pairs (rewrite sites) to the context $L \, \_ \, R$. For a single specified context $L \_ R$ contexts, we can filter out illicit centers by composing $\neg(\neg(\Gamma^*~ L)~ <~ \Gamma^*~ \cup~ \Gamma^*~ >~ \neg(R~ \Gamma^*)) \circ T_{base}$. However, a generalization of this formula grows exponentially with the number of context pairs $L_i \_ R_i$, which is why we resort to a method by Yli-Jyrä and Koskenniemi~\cite{yli2004compiling}. Define the center language
\[
C = < \, (\Gamma - >)^* \, >,
\]
which denotes a single bracketed rewrite site. The context restriction ensures that every occurrence of $C$ appears only in $L \, \_ \, R$.

Introduce a fresh auxiliary symbol $\cdot \notin \Gamma$, and define
\[
K = \widetilde{L} \, \cdot \, \Gamma^* \, \cdot \, \widetilde{R},
\]
where $\widetilde{L}$ and $\widetilde{R}$ are obtained from $L$ and $R$ by freely inserting symbols from $\{<,>\}$ between ordinary symbols. Intuitively, the two copies of $\cdot$ delimit a single site in the string. Thus $K$ is the language of strings in which the marked position is licensed by the left and right contexts.

We can now express the bad strings, namely those containing a dotted rewrite site that is not properly surrounded, as
\[
\Gamma^* \, \cdot \, C \, \cdot \, \Gamma^* \;-\; \Gamma^* K \Gamma^*.
\]
Deleting $\cdot$ then projects away the temporary marker, and negation yields the admissible candidates:

\begin{equation}
L_{context} = \neg h_{\{\cdot\}}(\Gamma^* \, \cdot \, C \, \cdot \, \Gamma^* - \Gamma^* K \Gamma^*)
\label{eqn:context}
\end{equation}

Here, $h_{\{\cdot\}}$ is the homomorphism deleting $\cdot$. With this, every rewrite site must satisfy the context test, and the two copies of $\cdot$ ensure that the same site is quantified in the two subexpressions \cite{hulden2009regular}.

Compose this with $T_{base}$ to get a rule transducer that rewrites relevant input strings nondeterministically (i.e.\ optionally), but only in the correct context:

\begin{equation}
T_{\rm optional\_rewrite} = L_{context} \circ T_{base}
\label{eqn:optionalrewrite}
\end{equation}

\subsection{Obligatory Rewriting}

The above produces a transducer that models optional rewriting in the correct contexts, since we have not yet introduced a mechanism that forces a legal rewrite to apply. To enforce obligatory rewriting, define a worsening transducer that removes at least one bracket pair, setting up the partial order:

\begin{equation}
W_{\rm obl} = \Gamma^* \underbrace{( <:\!\epsilon~ \Sigma^*~ >:\!\epsilon )}_{\text{Remove bracket pair}} \underbrace{( <:\!\epsilon~ \Sigma^* >:\!\epsilon \cup \Gamma)^*}_{\text{\parbox{3cm}{Remove more or repeat}}}
\end{equation}

\begin{figure}
\begin{center}
\begin{adjustbox}{scale=0.9}
\begin{tikzpicture}[scale=0.2]
\tikzset{
    state/.style={circle, draw=black, fill=gray!30, inner sep=0pt},
    final_inner/.style={circle, draw=black, fill=gray!30, inner sep=0pt}
}

\draw [black] (7,-31.5) -- (9,-31.5);
\fill [black] (9,-31.5) -- (8.2,-31) -- (8.2,-32);

\draw [black, fill=gray!30] (12,-31.5) circle (3);
\draw (12,-31.5) node {$0$};

\draw [black, fill=gray!30] (23.4,-31.5) circle (3);
\draw (23.4,-31.5) node {$1$};

\draw [black] (35.2,-31.5) circle (3);
\draw [black, fill=gray!30] (35.2,-31.5) circle (2.4);
\draw (35.2,-31.5) node {$2$};

\draw [black] (10.677,-28.82) arc (234:-54:2.25);
\draw (12,-24.25) node [above] {$.,\mbox{ }<:<,\mbox{ }>:>$};
\fill [black] (13.32,-28.82) -- (14.2,-28.47) -- (13.39,-27.88);

\draw [black] (15,-31.5) -- (20.4,-31.5);
\fill [black] (20.4,-31.5) -- (19.6,-31) -- (19.6,-32);
\draw (17.7,-31) node [above] {$<:\!\epsilon$};

\draw [black] (22.077,-28.82) arc (234:-54:2.25);
\draw (23.4,-24.25) node [above] {$.$};
\fill [black] (24.72,-28.82) -- (25.6,-28.47) -- (24.79,-27.88);

\draw [black] (26.181,-30.395) arc (104.68271:75.31729:12.305);
\fill [black] (32.42,-30.4) -- (31.77,-29.71) -- (31.52,-30.68);
\draw (29.3,-29.49) node [above] {$>:\!\epsilon$};

\draw [black] (32.577,-32.931) arc (-70.24202:-109.75798:9.694);
\fill [black] (26.02,-32.93) -- (26.61,-33.67) -- (26.94,-32.73);
\draw (29.3,-34) node [below] {$<:\!\epsilon$};

\draw [black] (33.877,-28.82) arc (234:-54:2.25);
\draw (35.2,-24.25) node [above] {$.,\mbox{ }<:<,\mbox{ }>:>$};
\fill [black] (36.52,-28.82) -- (37.4,-28.47) -- (36.59,-27.88);

\end{tikzpicture}
\end{adjustbox}
\end{center}

\caption{The worsening transducer $W_{obl}$ which nondeterministically removes at least one bracket pair. The symbol {\bf .} denotes PyFoma's catch-all transition label: in an FST, it repeats any symbol not explicitly listed in the transducer's current alphabet.}
\label{fig:wobl}

\end{figure}
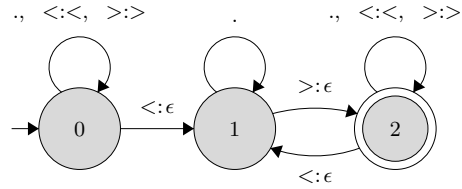

This maps, e.g., $\# \, a \, < \, a\, b \, > \, a \, \#$ to $\# \, a \, a \, b \, a \, \#$, undoing a rewrite site (Figure \ref{fig:wobl} shows the transducer). We now filter suboptimal candidates as in Section \ref{sec:worsening}, producing our final rule transducer:

\begin{equation}
T_{rule} = \neg \text{range}(\text{dom}(T_{\rm optional\_rewrite}) \circ W_{\text{obl}}) \circ T_{\rm optional\_rewrite}
\end{equation}

This retains candidates from which no additional legal non-overlapping rewrite site can be obtained by adding bracket pairs. As a final step, we remove the auxiliary symbols by applying the homomorphism $\{<,>,\#\} \mapsto \epsilon$.

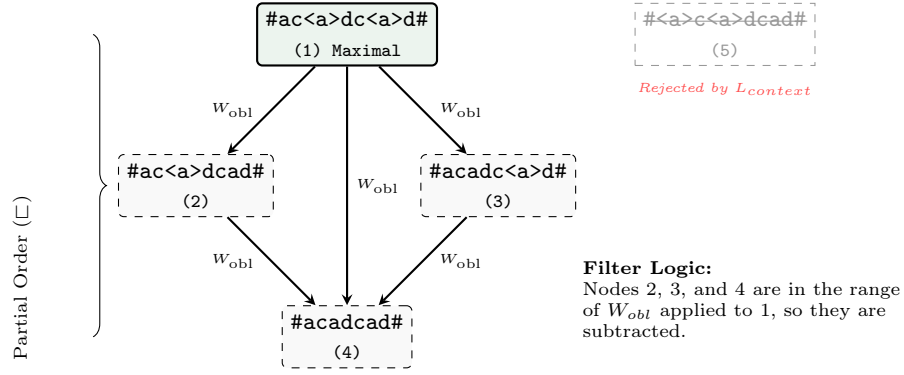
\begin{figure}

    \begin{tikzpicture}[node distance=2cm, auto, >=stealth]
    \small
    \ttfamily
    
    \node (n1) at (0, 4) [draw, thick, fill=green!10, rounded corners=2pt, align=center] 
    {\#ac<a>dc<a>d\#  \\  \scriptsize (1) \textbf{Maximal}};
        
    \node (n2) at (-2, 2) [draw, dashed, fill=gray!5, rounded corners=2pt, align=center] 
        {\#ac<a>dcad\# \\ \scriptsize (2)};
        
    \node (n3) at (2, 2) [draw, dashed, fill=gray!5, rounded corners=2pt, align=center] 
        {\#acadc<a>d\# \\ \scriptsize (3)};
        
    \node (n4) at (0, 0) [draw, dashed, fill=gray!5, rounded corners=2pt, align=center] 
        {\#acadcad\# \\ \scriptsize (4)};

    \node (n5) at (5, 4) [draw=red!50, color=gray!80, dashed, align=center] 
        {\st{\#<a>c<a>dcad\#} \\ \scriptsize (5)};
    \node[below=0.1cm of n5, font=\itshape\rmfamily\tiny, color=red!70] 
        {Rejected by $L_{context}$};

    \draw[->, thick] (n1) -- (n2) node[midway, left=0.1cm, font=\rmfamily\tiny] {$W_{\rm obl}$};
    \draw[->, thick] (n1) -- (n3) node[midway, right=0.1cm, font=\rmfamily\tiny] {$W_{\rm obl}$};
    \draw[->, thick] (n2) -- (n4) node[midway, left=0.1cm, font=\rmfamily\tiny] 
    {$W_{\rm obl}$};
    \draw[->, thick] (n3) -- (n4) node[midway, right=0.1cm, font=\rmfamily\tiny]
    {$W_{\rm obl}$};
        \draw[->, thick] (n1) -- (n4) node[midway, right=0.0cm, font=\rmfamily\tiny]
    {$W_{\rm obl}$};
    \draw [decorate, decoration={brace,amplitude=5pt,mirror,raise=4pt}, yshift=0pt]
    (-3.5,0) -- (-3.5,4) node [black,midway,xshift=-0.8cm, rotate=90, font=\rmfamily\scriptsize] {Partial Order ($\sqsubset$)};

    \node[anchor=west, font=\rmfamily\scriptsize, align=left] at (3, 0.5) {
        \textbf{Filter Logic:}\\
        Nodes 2, 3, and 4 are in the range \\
        of $W_{obl}$ applied to 1, so they are \\
        subtracted.
    };

\end{tikzpicture}
\caption{Some candidates generated by $T_{base}$ for $a \to b / c \_ d$ on input \texttt{\#acadcad\#}. Candidate 5 contains a rewritten site outside the context $c \_ d$ and is removed by $L_{context}$. Candidate 1 has competitors 2, 3, and 4 obtained by deleting one or both bracket pairs, so only 1 survives obligatory filtering. Obligatory worsening compares bracketed candidate strings on the input side; only those representations are shown here.}
\label{fig:obl_candidates}
\end{figure}

The logic of the construction is simple. $T_{base}$ generates all candidate bracketings and replacements. $L_{context}$ removes those with a bracketed site in an illicit context, leaving the legal optional rewrites. $W_{\text{obl}}$ then makes one legal candidate worse than another if it is obtained by removing one or more bracket pairs, i.e., by undoing one or more legal rewrites. Hence $\neg \text{range}(\text{dom}(T_{\rm optional\_rewrite}) \circ W_{\text{obl}}) \circ T_{\rm optional\_rewrite}$ keeps exactly those legal candidates from which no legal rewrite has been omitted. Figure~\ref{fig:obl_candidates} illustrates this for the rule $a \to b / c \_ d$.

A useful property of the obligatory construction is that the resulting rule transducer is total on $\Sigma^*$. For every input string, $T_{base}$ admits the candidate in which no rewrite is applied, so all symbols are copied identically and no rewrite brackets are inserted. This candidate trivially survives context filtering, since $L_{context}$ excludes only strings containing a bracketed site in an illicit context. Moreover, on the candidates that survive this step, $W_{\text{obl}}$ is acyclic, since it worsens a candidate only by removing one or more bracket pairs. For any fixed input, only finitely many bracketings are possible, so at least one surviving candidate is not worsened by any other, and the final filtering step therefore cannot eliminate them all.

The above gives the basic construction for a rule $A \to B / L~ \_~ R$. We now briefly consider extensions to this basic case.

\subsection{Multiple Contexts}

For multiple alternative rewrite contexts
\[
L_1 \, \_ \, R_1, \dots, L_n \, \_ \, R_n,
\]
the same dotted-position construction extends directly by union. Define
\[
K = \bigcup_{i=1}^n \widetilde{L_i} \, \cdot \, \Gamma^* \, \cdot \, \widetilde{R_i},
\]
where each $\widetilde{L_i}$ and $\widetilde{R_i}$ is obtained from $L_i$ and $R_i$ by freely inserting symbols from $\{<,>\}$ between ordinary symbols. Thus $K$ is the language of strings in which the $\cdot$-marked span is surrounded by at least one allowed context pair. In other words, disjunction of contexts is handled entirely by union at the level of the dotted context language.

Using the same center language
\[
C = < \, (\Gamma - >)^* \, >,
\]
we obtain
\begin{equation}
L_{context} = \neg h_{\{\cdot\}}(\Gamma^* \, \cdot \, C \, \cdot \, \Gamma^* - \Gamma^* K \Gamma^*)
\end{equation}

This ensures that every rewrite site belongs to at least one valid context pair \cite{yli2004compiling}. The worsening transducer $W_{obl}$ is unchanged.

\subsection{Directed Rewriting}

To control ambiguity (e.g., $a^+b^+c^+ \to x$ on input $aabbcc$, where the four matching spans $abbc$, $abbcc$, $aabbc$, and $aabbcc$ produce $\{axc, ax, xc, x\}$), define further worsening transducers:

\begin{itemize}
\item {\bf Leftmost}: $W_{\text{left}}$, shifting opening brackets rightward, e.g., $< aabbcc > \to a < abbcc >$.
\item {\bf Longest}: $W_{\text{long}}$, shortening spans, e.g., $< aabbcc > \to < aabbc > c$.
\item {\bf Shortest}: $W_{\text{short}}$, lengthening spans, e.g., $< aabbc > c \to < aabbcc >$.
\end{itemize}

See Section \ref{ref:directed_worseners} for the exact constructions of the worsening transducers above.

When several criteria are active, we combine their worsening relations by union and construct the corresponding filter exactly as in the basic case. Thus a candidate is excluded whenever some competing candidate for the same input is better under \emph{any} one of the active criteria. For example, to require rewrites to be obligatory, leftmost, and longest, we construct:
\begin{equation}
T_{rule} = \neg\text{range}(\text{dom}(T_{\rm optional\_rewrite}) \circ (W_{\text{obl}} \cup W_{\text{left}} \cup W_{\text{long}})) \circ T_{\rm optional\_rewrite}
\end{equation}
So omitted obligatory rewrites, non-leftmost choices, and non-longest choices are each sufficient for exclusion. For example, applying $a^+ \to x / a \, \_ \, a$ to $aaaaa$ with all three criteria active yields $axa$ as the only possible output. In PyFoma, this is expressed as:

\begin{lstlisting}
$^rewrite(a+:x  / a _ a, leftmost = True, longest = True)
\end{lstlisting}

\subsection{Arbitrary Transductions}
\label{sec:arb}

The input $T$ can be any FST, not necessarily a cross-product of regular languages $A\!:\!B$. In PyFoma, the basic interface is \lstinline|$^rewrite(T / L1 _ R1, ..., Ln _ Rn)|.

A word-final devoicing rule for stops (b,d,g) is:

\begin{lstlisting}
$^rewrite(b:p|g:k|d:t / _ #)
\end{lstlisting}

This allows arbitrary transductions without introducing additional rule types.

\subsection{Markup Transducers}

Markup transducers, common in tools such as {\it xfst} and {\it foma} \cite{beesley2003finite}, insert material on either or both sides of some target string. Such transducers are commonly used for HTML-style markup \cite{beesley2003finite}. A simple example is wrapping vowels with tags while leaving the vowel itself unchanged:

\begin{lstlisting}[basicstyle=\fontsize{8}{10}\selectfont\ttfamily]
$^rewrite('':'<VOWEL>' [aeiou] '':'</VOWEL>')
\end{lstlisting}





Other toolkits often have a special rule type to perform this kind of insertion. Because our rules are not restricted to cross-products, we need no special treatment for such transductions.

\subsection{Parallel Rewriting}

Parallel rewriting, as in Kempe and Karttunen \cite{kempe1996parallel}, applies multiple rules simultaneously. Consider the pair of rules
\[
ab \to x / \_ ba, \qquad ba \to y / ab \_
\]
on input \texttt{abba} (i.e., \texttt{\#abba\#}). One candidate generated by an indexed extension of $T_{\rm base}$ marks the two non-overlapping rewrite sites with indexed brackets:

\begin{center}
\texttt{\#<}\(_{1}\)\colorbox{rewritehl}{\texttt{ab}}\texttt{>}\(_{1}\)\texttt{<}\(_{2}\)\colorbox{rewritehl}{\texttt{ba}}\texttt{>}\(_{2}\)\texttt{\#}
\end{center}

Context restriction (via the dotted-position method) and obligatory worsening are then enforced separately for each indexed bracket pair \((<_{i},\, >_{i})\). The same typed-bracket mechanism extends directly to any number of parallel rules and to the different rewrite strategies discussed above (obligatory, leftmost, longest, etc.); in the implementation, the indices and auxiliary brackets are projected away in the final step exactly as in the single-rule case, yielding, in the present example:

\begin{center}
\texttt{\colorbox{lblue}{xy}}
\end{center}

\subsection{Weights}

PyFoma supports weights in the tropical semiring \cite{pin1998}, and weighted compilation fits naturally into the same construction because worseners never compare weights. For example, word-final devoicing with cost 1.0:

\begin{lstlisting}[basicstyle=\fontsize{8}{10}\selectfont\ttfamily]
$^rewrite((b:p | g:k | d:t)<1.0> / _ #)
\end{lstlisting}

Allowing non-rewriting with a different cost:

\begin{lstlisting}[basicstyle=\fontsize{8}{10}\selectfont\ttfamily]
$^rewrite(((b:p | g:k | d:t)<1.0> | [bdg]<0.0>) / _ #)
\end{lstlisting}

No special treatment is required: weights enter only through the underlying transduction $T$, while the later stages merely filter candidates, so the weights of surviving candidates are exactly those assigned by $T$.

\subsection{Epsilons in Left-Hand Sides}

For many applications, rewrite rules whose left-hand side contains $\epsilon$ require a more restricted interpretation. A rule such as $\epsilon \to x$ or $b^* \to x$ would otherwise license zero or arbitrarily many insertions at the same position. Often, the desired behavior is to allow at most one insertion site at each boundary between adjacent positions, including word edges, so that input \texttt{aa} yields the single output \texttt{xaxax} rather than arbitrarily many insertions at the same point.

PyFoma adopts by default the bounded interpretation just described. This is enforced by an initial filter excluding candidate strings that contain consecutive empty bracket pairs $<><>$.

The transducer $T_{base}$ is restricted on its input side to exclude consecutive bracket pairs using:

\begin{equation}
L_{\rm dotted} = \Gamma^* - (\Gamma^* \, < \, > \, < \, > \, \Gamma^*)
\end{equation}

Thus, we modify the earlier $T_{base}$; $T_{base} = L_{\rm dotted} \circ \# \, (\Sigma~ \cup < \, T \, >)^* \, \#$.\footnote{Such rules were called {\bf dotted} rules in the xfst-toolkit \cite{beesley2003finite} and we inherit the name.}

\subsection{Spreading and Harmony Rules}
\label{sec:output_side}

Karttunen \cite{karttunen1995replace} noted that one can achieve different behavior of rewrite rules depending on whether the contextual constraints $L_i~ \_~ R_i$ are assumed to hold on the input side or on the output side.  The traditional assumption, and the default adopted here, is that they hold on the input side.  However, by swapping the order of composition in (\ref{eqn:optionalrewrite}), we can constrain the contexts to hold on the output side:

\begin{equation}
T_{\rm optional\_rewrite} = T_{base} \circ L_{context}
\end{equation}

A minimal example of the different behavior is seen in a rule such as 

$$ab \to x /~ ab~ \_~ a$$ 

\noindent and the input $abababa$, where constraining the context on the input yields one output, $abxxa$, whereas constraining the context on the output side produces two different outputs, $\{ababxa,abxaba\}$\footnote{\cite{karttunen1995replace} actually makes a 4-way distinction based on whether the left and right contexts hold on the input or output side.} (see Table~\ref{tab:inputoutput}).

\begin{table}[ht]
\centering
\renewcommand{\arraystretch}{1.15}
\setlength{\tabcolsep}{8pt}
\begin{tabular}{l l l}
\toprule
& \textbf{Input-side context} & \textbf{Output-side context} \\
\midrule
Input &
\texttt{a b \colorbox{rewritehl}{a b} \colorbox{rewritehl}{a b} a} &
\texttt{a b \colorbox{rewritehl}{a b} \colorbox{rewritehl}{a b} a} \\
\cdashline{1-3}
Output(s) &
\texttt{a b \hspace{0em} \colorbox{lblue}{x} \hspace{.5em} \colorbox{lblue}{x} \hspace{0em} a} &
\begin{tabular}[t]{@{}l@{}}
\texttt{a b \hspace{-.2em} a b \hspace{.3em} \colorbox{lblue}{x} \hspace{0.05em} a} \\
\texttt{a b \hspace{.4em}\colorbox{lblue}{x} \hspace{.5em} a \hspace{-.5em} b \hspace{-.2em} a}
\end{tabular} \\
\bottomrule
\addlinespace[1.0ex]
\end{tabular}
\caption{The rule $ab \to x / ab \_ a$ on input \texttt{abababa}, with contexts checked on the input side vs.\ output side.}
\label{tab:inputoutput}
\end{table}

Although the input-side restriction often gives the expected behavior for ordinary rewrite rules, output-side restriction has an important use in computational phonology, related to vowel-harmony rules. This arises when a vowel acquires its quality from a preceding vowel in the word. In Finnish, certain suffix vowels are underspecified and acquire their surface quality from the preceding vowel \cite{karlsson2013finnish}. For example, the word {\bf kyl\"a} `village' has front vowels, \{{\bf y,\"a}\} in it, and any suffixes added to it should also receive front vowels. Commonly, these underspecified suffixes are denoted with capital letters (A = either a or \"a, O = either o or \"o).  A typical rule to handle the front-vowel case would rewrite A as \"a if the previous vowel is \"a or \"o or y, and O as \"o in the same circumstance.  With a rule like 

\begin{lstlisting}[
    basicstyle=\fontsize{8}{10}\selectfont\ttfamily,
    literate={ä}{{\"a}}1 {ö}{{\"o}}1 {Ä}{{\"A}}1 {Ö}{{\"O}}1
]
$^rewrite(A:ä | O:ö / (ä|ö|y) $nonvowel+ _ )
\end{lstlisting}

\noindent this can be achieved, but only if the contexts hold on the output side, which allows the vowel quality to `spread' rightward (see Table~\ref{tab:harmony}).

\vspace{-1ex}

\begin{table}[!ht]
\centering
\renewcommand{\arraystretch}{1.15}
\setlength{\tabcolsep}{8pt}
\begin{tabular}{l l l}
\toprule
& \textbf{Input-side context} & \textbf{Output-side context} \\
\midrule
Input &
\texttt{k y l ä + s s \colorbox{rewritehl}{A} k O} &
\texttt{k y l ä + s s \colorbox{rewritehl}{A} k \colorbox{rewritehl}{O}} \\
\cdashline{1-3}
Output(s) &
\texttt{k y l ä + s s \colorbox{lblue}{ä} k O} &
\texttt{k y l ä + s s \colorbox{lblue}{ä} k \colorbox{lblue}{ö}} \\
\bottomrule
\addlinespace[1.0ex]
\end{tabular}
\caption{A harmony rule whose context must be checked on the output side to allow the feature to spread rightward.}
\label{tab:harmony}
\end{table}

\vspace*{-8.5ex}

\section{Conclusion}
\label{sec:conclusion}

We have presented a rewrite compilation pattern built from three ingredients: candidate generation, context restriction, and preference filtering by worsening. The same pattern covers obligatory rewriting, multiple contexts, directed rewriting, arbitrary transductions, markup, parallel rules, and weights.

More broadly, rewrite compilation is only one application of worsening. Although the technique arose in finite-state phonology, it is better understood as a general finite-state design pattern for turning procedural-looking selection tasks into algebraic ones: generate a regular set of candidates, define a worsening relation, and filter away the non-optimal candidates. Besides its established use in finite-state Optimality Theory \cite{gerdemann-van-noord-2000-approximation,gerdemann2012practical}, this perspective points to a wider role for worsening in extracting preferred subsets of regular languages and relations---for example shortest, longest, or otherwise minimal representatives---and in encoding complex preference orders declaratively within the regular calculus itself.

\begin{credits}

\subsubsection{\discintname}
The authors have no competing interests to declare that are
relevant to the content of this article.
\end{credits}

\appendix
\renewcommand{\theHsection}{app.\thesection}
\section{PyFoma Implementation}
\label{ref:pyfoma_implementation}

The following PyFoma code illustrates the rewrite rule compilation, using {\tt <} and {\tt >} for brackets and {\tt \#} for word boundaries. It relies on the built-in {\tt \$\^{}restrict()} for context restriction. The code is a simplified adaptation from PyFoma's actual {\tt \$\^{}rewrite()} functionality, which handles multiple contexts and modalities. As in {\it foma} and {\it xfst}, the symbol {\bf .} has two related interpretations: in regular expressions it denotes any symbol, while in compiled FSTs it functions as a catch-all transition for symbols not explicitly present in the machine's current alphabet.

\begin{lstlisting}[caption={PyFoma implementation of a rewrite rule compilation $A \to B~ /~ L~ \_~ R$},label={lst:rewrite}, basicstyle=\fontsize{8}{10}\selectfont\ttfamily]
from pyfoma import re
t = {}
t['A'] = re("a")       # A, B, L, R
t['B'] = re("b")       # can be
t['L'] = re("c")       # arbitrary regular languages
t['R'] = re("d")       # use # for edges in contexts

t['T']         = re("$A:$B<1.0>", t) # Cross-product with weight
t['sigma']     = re("[^<>#]")
t['base']      = re("# (< $T > | $sigma)* #", t)
t['leftc']     = re("$^ignore($L, [<>])", t)
t['rightc']    = re("$^ignore($R, [<>])", t)
t['context']   = re("$^restrict(< [^>]* > / $leftc _ $rightc)", t)
t['optrewr']   = re("$context @ $base", t)
t['remrewr']   = re(" <:'' [^>]* >:'' ")
t['oblworsen'] = re(".* $remrewr (. | $remrewr)*", t)
t['badrewr']   = re("$^output($^input($optrewr) @ $oblworsen)", t)
t['oblrewr']   = re("~$badrewr @ $optrewr", t)
t['rule']      = t['oblrewr'].map_labels({'<':'', '>':'', '#':''})\
                  .epsilon_remove().determinize().minimize_as_dfa()
\end{lstlisting}

\begin{figure}[htbp]
\begin{center}
\begin{adjustbox}{width=0.70\textwidth}
\begin{tikzpicture}[scale=0.2]
\tikzset{
    state/.style={circle, draw=black, fill=gray!30, inner sep=0pt},
}

\draw [black] (6.5,-31.1) -- (8.5,-31.1);
\fill [black] (8.5,-31.1) -- (7.7,-30.6) -- (7.7,-31.6);

\draw [black] (11.5,-31.1) circle (3);
\draw [black, fill=gray!30] (11.5,-31.1) circle (2.4);
\draw (11.5,-31.1) node {$0/0$};

\draw [black] (25.9,-31.1) circle (3);
\draw [black, fill=gray!30] (25.9,-31.1) circle (2.4);
\draw (25.9,-31.1) node {$1/0$};

\draw [black] (40.6,-31.1) circle (3);
\draw [black, fill=gray!30] (40.6,-31.1) circle (2.4);
\draw (40.6,-31.1) node {$2/0$};

\draw [black, fill=gray!30] (25.9,-42.6) circle (3);
\draw (25.9,-42.6) node {$3$};

\draw [black] (10.177,-28.42) arc (234:-54:2.25);
\draw (11.5,-23.85) node [above, yshift = -4pt] {$./0,\mbox{ }a\!:\!a/0,\mbox{ }b\!:\!b/0,\mbox{ }d\!:\!d/0$};
\fill [black] (12.82,-28.42) -- (13.7,-28.07) -- (12.89,-27.48);

\draw [black] (14.265,-29.948) arc (106.96687:73.03313:15.199);
\fill [black] (23.14,-29.95) -- (22.52,-29.24) -- (22.22,-30.19);
\draw (18.7,-28.79) node [above, yshift = -4pt] {$c\!:\!c/0$};

\draw [black] (23.596,-33.002) arc (-59.39103:-120.60897:9.616);
\fill [black] (13.8,-33) -- (14.24,-33.84) -- (14.75,-32.98);
\draw (18.7,-34.84) node [below, yshift = 3pt] {$./0,\mbox{ }b\!:\!b/0,\mbox{ }d\!:\!d/0$};

\draw [black] (28.716,-30.076) arc (105.05956:74.94044:17.451);
\fill [black] (37.78,-30.08) -- (37.14,-29.39) -- (36.88,-30.35);
\draw (33.25,-28.98) node [above, yshift = -4pt] {$a\!:\!a/0$};

\draw [black] (24.577,-28.42) arc (234:-54:2.25);
\draw (25.9,-23.85) node [above, yshift=-4pt] {$c\!:\!c/0$};
\fill [black] (27.22,-28.42) -- (28.1,-28.07) -- (27.29,-27.48);

\draw [black] (38.152,-32.817) arc (-62.93401:-117.06599:10.773);
\fill [black] (28.35,-32.82) -- (28.83,-33.63) -- (29.29,-32.74);
\draw (33.25,-34.5) node [below, yshift=4pt, xshift = 2pt] {$c\!:\!c/0$};

\draw [black] (25.9,-34.1) -- (25.9,-39.6);
\fill [black] (25.9,-39.6) -- (26.4,-38.8) -- (25.4,-38.8);
\draw (26.4,-36.85) node [right, xshift=-4pt, yshift = 2pt] {$a\!:\!b/1$};

\draw [black] (23.019,-43.397) arc (-82.79713:-174.42538:10.404);
\fill [black] (11.36,-34.09) -- (10.94,-34.93) -- (11.94,-34.83);
\draw (12.66,-41.7) node [above, xshift = 35pt, yshift = -6pt] {$d\!:\!d/0$};

\draw [black] (40.638,-34.095) arc (-5.1359:-174.8641:14.647);
\fill [black] (11.46,-34.09) -- (11.04,-34.94) -- (12.03,-34.85);
\draw (26.05,-47.93) node [below, yshift = 4pt] {$./0,\mbox{ }a\!:\!a/0,b\!:\!b/0$};
\end{tikzpicture}
\end{adjustbox}
\vspace{-22pt}
\end{center}
\caption{Resulting transducer compiled from Listing \ref{lst:rewrite}.}
\label{fig:abcd}
\end{figure}
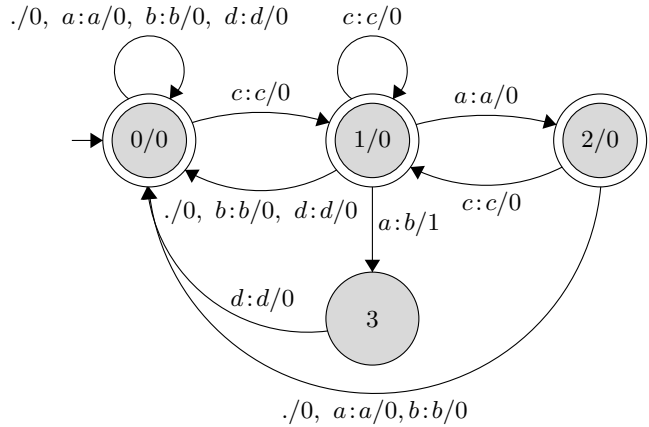

\section{Directed Rewriting Worseners}
\label{ref:directed_worseners}

For compactness, we use postfix $?$ for optional material, so that $X?$ abbreviates $(X \cup \epsilon)$.

\begin{equation}
\begin{aligned}
W_{\rm short} &= 
    \Gamma^*~
    \underbrace{< \Sigma^*~ >:\!\epsilon}_{\substack{\text{keep span,}\\\text{delete old } >}}
    \underbrace{\Sigma^+~ \epsilon:>}_{\substack{\text{extend span}\\\text{rightward}}}
    ~\Gamma^* \\[1ex]
W_{\rm long} &= 
    \Gamma^*~
    \underbrace{<~\Sigma^+~\epsilon:(>~<?)~\Sigma}_{\substack{\text{close earlier,}\\\text{optionally reopen}}}
    \underbrace{\big((< \cup >):\epsilon \cup \epsilon:(< \cup >) \cup \Sigma\big)^*}_{\substack{\text{copy remainder and}\\\text{clean up old brackets}}}
    ~\Gamma^* \\[1ex]
W_{\rm left} &=
    \Gamma^*~
    \underbrace{<:\epsilon~ \Sigma^+~ \epsilon:<~ \Sigma^*}_{\substack{\text{delete old } <, \text{ then} \\ \text{re-insert farther right}}}
    \left(
        \underbrace{\epsilon:>~ \Sigma^+~ >:\epsilon}_{\substack{\text{new match ends} \\ \text{before old match}}}~
        \cup
        \underbrace{>:\epsilon~ \Sigma^*~ \epsilon:>}_{\substack{\text{new match ends at} \\ \text{or after old match}}}
    \right) ~\Gamma^*
\end{aligned}
\end{equation}

\begin{figure}
\begin{center}
\begin{adjustbox}{scale=0.9}
\begin{tikzpicture}[scale=0.2]
\tikzset{
    state/.style={circle, draw=black, fill=gray!30, inner sep=0pt},
}

\draw [black] (6.5,-31.1) -- (8.5,-31.1);
\fill [black] (8.5,-31.1) -- (7.7,-30.6) -- (7.7,-31.6);

\foreach \pos/\name in {{(11.5,-31.1)/0}, {(22.9,-31.1)/1}, {(34.6,-31.1)/2}, {(46.2,-31.1)/3}} {
    \draw [black, fill=gray!30] \pos circle (3);
    \draw \pos node {$\name$};
}

\draw [black] (58.1,-31.1) circle (3);
\draw [black, fill=gray!30] (58.1,-31.1) circle (2.4);
\draw (58.1,-31.1) node {$4$};

\draw [black] (10.177,-28.42) arc (234:-54:2.25);
\draw (11.5,-23.85) node [above] {$.,\mbox{ }\#\!:\!\#,\mbox{}>:>$};
\fill [black] (12.82,-28.42) -- (13.7,-28.07) -- (12.89,-27.48);

\draw [black] (14.066,-29.574) arc (110.94834:69.05166:8.767);
\fill [black] (20.33,-29.57) -- (19.77,-28.82) -- (19.41,-29.75);
\draw (17.2,-28.49) node [above] {$<:<$};

\draw [black] (20.225,-32.432) arc (-72.21277:-107.78723:9.901);
\fill [black] (14.18,-32.43) -- (14.78,-33.15) -- (15.09,-32.2);
\draw (17.2,-33.41) node [below] {$\#\!:\!\#,\mbox{}>:>$};

\draw [black] (21.577,-28.42) arc (234:-54:2.25);
\draw (22.9,-23.85) node [above] {$.,\mbox{}<:<$};
\fill [black] (24.22,-28.42) -- (25.1,-28.07) -- (24.29,-27.48);

\draw [black] (25.9,-31.1) -- (31.6,-31.1);
\fill [black] (31.6,-31.1) -- (30.8,-30.6) -- (30.8,-31.6);
\draw (28.75,-30.6) node [above] {$>:\!\epsilon$};

\draw [black] (37.6,-31.1) -- (43.2,-31.1);
\fill [black] (43.2,-31.1) -- (42.4,-30.6) -- (42.4,-31.6);
\draw (40.4,-30.6) node [above] {$.$};

\draw [black] (44.877,-28.42) arc (234:-54:2.25);
\draw (46.2,-23.85) node [above] {$.$};
\fill [black] (47.52,-28.42) -- (48.4,-28.07) -- (47.59,-27.48);

\draw [black] (49.2,-31.1) -- (55.1,-31.1);
\fill [black] (55.1,-31.1) -- (54.3,-30.6) -- (54.3,-31.6);
\draw (52.15,-30.6) node [above] {$\epsilon\!:>$};

\draw [black] (56.777,-28.42) arc (234:-54:2.25);
\draw (58.1,-23.85) node [above] {$.,\mbox{ }\#\!:\!\#,\mbox{ }<:<,\mbox{}>:>$};
\fill [black] (59.42,-28.42) -- (60.3,-28.07) -- (59.49,-27.48);

\end{tikzpicture}
\end{adjustbox}
\vspace{-15pt}
\end{center}
\caption{The worsener $W_{\rm short}$. The symbol {\bf .} denotes PyFoma's catch-all transition label: in an FST, it repeats any symbol not explicitly listed in the transducer's current alphabet.}
\label{fig:wshort}
\end{figure}
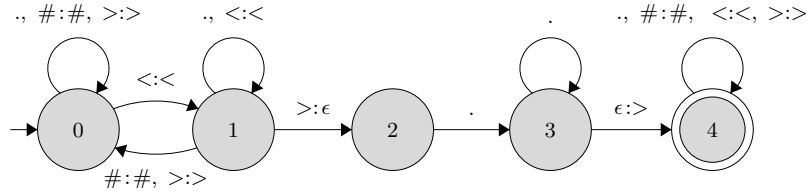

$W_{\rm short}$ moves a closing bracket rightward, making a rewrite span less short. $W_{\rm long}$ moves a closing bracket leftward, optionally reopening immediately so that later bracket structure can still be preserved. Finally, $W_{\rm left}$ moves the opening bracket rightward. Its initial move is the mirror image of $W_{\rm short}$ (Figure \ref{fig:wshort}), but the closing bracket cannot simply be left untouched: a later-starting competing span may end earlier, at the same point, or later. The two alternatives therefore cover the possible placements of the new closing bracket relative to the old one. In each case, the worsener relates a preferred candidate to a competing candidate that is worse with respect to the corresponding criterion.

\noindent Examples of single applications to bracketed candidate strings:

\bigskip

\begin{itemize}
  \item $W_{\rm short}$: $\langle \texttt{aabbc} \rangle \texttt{c} \mapsto \langle \texttt{aabbcc} \rangle$ \quad (moves $>$ right)
  \item $W_{\rm long}$:  $\langle \texttt{aabbcc} \rangle \mapsto \langle \texttt{aabbc} \rangle \texttt{c}$ \quad (moves $>$ left)
  \item $W_{\rm left}$: $\langle \texttt{aabbcc} \rangle \mapsto \texttt{a} \langle \texttt{abbcc} \rangle$ \quad (moves $<$ right)
\end{itemize}

\bibliographystyle{splncs04}
\bibliography{fstbib}

@book{beesley2003finite,
  author = {Beesley, Kenneth R. and Karttunen, Lauri},
  title = {Finite-State Morphology},
  year = {2003},
  publisher = {CSLI Publications},
  address = {Stanford}
}

@book{chomsky1968sound,
  author = {Chomsky, Noam and Halle, Morris},
  title = {The Sound Pattern of English},
  year = {1968},
  publisher = {Harper and Row},
  address = {New York}
}

@inproceedings{gerdemann-van-noord-2000-approximation,
    title = "Approximation and Exactness in Finite State {O}ptimality {T}heory",
    author = "Gerdemann, Dale  and
      van Noord, Gertjan",
    editor = "Eisner, Jason  and
      Karttunen, Lauri  and
      Th{\`e}riault, Alain",
    booktitle = "Proceedings of the Fifth Workshop of the {ACL} Special Interest Group in Computational Phonology",
    month = aug,
    year = "2000",
    address = "Centre Universitaire, Luxembourg",
    publisher = "International Committee on Computational Linguistics",
    url = "https://aclanthology.org/W00-1804/",
    pages = "34--45"
}

@inproceedings{gerdemann-2009-mix,
    title = "Mix and Match Replacement Rules",
    author = "Gerdemann, Dale",
    editor = "Bel, N{\'u}ria  and
      Hinrichs, Erhard  and
      Osenova, Petya  and
      Simov, Kiril",
    booktitle = "Proceedings of the Workshop on Adaptation of Language Resources and Technology to New Domains",
    month = sep,
    year = "2009",
    address = "Borovets, Bulgaria",
    publisher = "Association for Computational Linguistics",
    url = "https://aclanthology.org/W09-4106/",
    pages = "39--47"
}

@inproceedings{gerdemann2012practical,
    title = "Practical Finite State {O}ptimality {T}heory",
    author = "Gerdemann, Dale  and
      Hulden, Mans",
    editor = "Alegria, I{\~n}aki  and
      Hulden, Mans",
    booktitle = "Proceedings of the 10th International Workshop on Finite State Methods and Natural Language Processing",
    month = jul,
    year = "2012",
    address = "Donostia{--}San Sebasti{\'a}n",
    publisher = "Association for Computational Linguistics",
    url = "https://aclanthology.org/W12-6202/",
    pages = "10--19"
}

@book{gorman2022finite,
  title={Finite-state text processing},
  author={Gorman, Kyle and Sproat, Richard},
  year={2022},
  publisher={Springer Nature}
}

@inproceedings{hulden-etal-2024-pyfoma,
    title = "{P}y{F}oma: a {P}ython finite-state compiler module",
    author = "Hulden, Mans  and
      Ginn, Michael  and
      Silfverberg, Miikka  and
      Hammond, Michael",
    editor = "Cao, Yixin  and
      Feng, Yang  and
      Xiong, Deyi",
    booktitle = "Proceedings of the 62nd Annual Meeting of the Association for Computational Linguistics (Volume 3: System Demonstrations)",
    month = aug,
    year = "2024",
    address = "Bangkok, Thailand",
    publisher = "Association for Computational Linguistics",
    url = "https://aclanthology.org/2024.acl-demos.24/",
    doi = "10.18653/v1/2024.acl-demos.24",
    pages = "258--265"
}

@phdthesis{hulden2009finite,
  author = {Hulden, Mans},
  title = {Finite-State Machine Construction Methods and Algorithms for Phonology and Morphology},
  year = {2009},
  school = {University of Arizona}
}

@inproceedings{hulden2009foma,
  author = {Hulden, Mans},
  title = {Foma: A Finite-State Compiler and Library},
  booktitle = {Proceedings of the 12th Conference of the European Chapter of the Association for Computational Linguistics: Demonstrations Session},
  year = {2009},
  pages = {29--32},
  publisher = {Association for Computational Linguistics}
}

@incollection{hulden2009regular,
  author = {Hulden, Mans},
  title = {Regular Expressions and Predicate Logic in Finite-State Language Processing},
  booktitle = {Finite-State Methods and Natural Language Processing: Post-Proceedings of the 7th International Workshop FSMNLP 2008},
  editor = {Piskorski, Jakub and Watson, Bruce and Yli-Jyr{\"a}, Anssi},
  year = {2009},
  pages = {82--89},
  publisher = {IOS Press}
}

@book{johnson1972formal,
  author = {Johnson, C. Douglas},
  title = {Formal Aspects of Phonological Description},
  year = {1972},
  publisher = {Mouton},
  address = {The Hague}
}

@inproceedings{beesley-2012-kleene,
    title = "{K}leene, a Free and Open-Source Language for Finite-State Programming",
    author = "Beesley, Kenneth R.",
    editor = "Alegria, I{\~n}aki  and
      Hulden, Mans",
    booktitle = "Proceedings of the 10th International Workshop on Finite State Methods and Natural Language Processing",
    month = jul,
    year = "2012",
    address = "Donostia{--}San Sebasti{\'a}n",
    publisher = "Association for Computational Linguistics",
    url = "https://aclanthology.org/W12-6209/",
    pages = "50--54"
}

@article{kaplan1994regular,
  author = {Kaplan, Ronald M. and Kay, Martin},
  title = {Regular Models of Phonological Rule Systems},
  journal = {Computational Linguistics},
  volume = {20},
  number = {3},
  year = {1994},
  pages = {331--378}
}

@book{karlsson2013finnish,
  title={Finnish: An essential grammar},
  author={Karlsson, Fred},
  year={2013},
  publisher={Routledge}
}

@inproceedings{karttunen1995replace,
  author = {Karttunen, Lauri},
  title = {The Replace Operator},
  booktitle = {Proceedings of the 33rd Annual Meeting of the Association for Computational Linguistics},
  year = {1995},
  pages = {16--23},
  publisher = {Association for Computational Linguistics},
  address = {Cambridge, MA}
}

@inproceedings{karttunen1996directed,
  author = {Karttunen, Lauri},
  title = {Directed Replacement},
  booktitle = {Proceedings of the 34th Annual Meeting of the Association for Computational Linguistics},
  year = {1996},
  pages = {108--115},
  publisher = {Association for Computational Linguistics},
  address = {Santa Cruz, CA}
}

@inproceedings{kempe1996parallel,
  author = {Kempe, Andr{\'e} and Karttunen, Lauri},
  title = {Parallel Replacement in Finite-State Calculus},
  booktitle = {Proceedings of the 16th International Conference on Computational Linguistics (COLING'96)},
  year = {1996},
  pages = {622--627},
  address = {Copenhagen, Denmark}
}

@inproceedings{mohri1996efficient,
  author = {Mohri, Mehryar and Sproat, Richard},
  title = {An Efficient Compiler for Weighted Rewrite Rules},
  booktitle = {Proceedings of the 34th Annual Meeting of the Association for Computational Linguistics},
  year = {1996},
  pages = {231--238},
  publisher = {Association for Computational Linguistics},
  address = {Santa Cruz, CA}
}

@incollection{pin1998,
  TITLE = {{Tropical Semirings}},
  AUTHOR = {Pin, Jean-Eric},
  URL = {https://hal.science/hal-00113779},
  BOOKTITLE = {{Idempotency (Bristol, 1994)}},
  EDITOR = {J. Gunawardena},
  PUBLISHER = {{Cambridge Univ. Press, Cambridge}},
  SERIES = {Publ. Newton Inst. 11},
  PAGES = {50-69},
  YEAR = {1998},
  HAL_ID = {hal-00113779},
  HAL_VERSION = {v1},
}

@phdthesis{vaillette2004logical,
  author = {Vaillette, Nathan},
  title = {Logical Specification of Finite-State Transductions for Natural Language Processing},
  year = {2004},
  school = {Ohio State University}
}

@InProceedings{yli2004compiling,
  author    = {Yli-Jyr{\"a}, Anssi and Koskenniemi, Kimmo},
  title     = {Compiling Contextual Restrictions on Strings into Finite-State Automata},
  booktitle = {The Eindhoven FASTAR Days, Proceedings},
  year      = {2004},
  editor    = {Cleophas, Loek and Watson, Bruce W.},
  publisher = {Technische Universiteit Eindhoven},
  address   = {Eindhoven, The Netherlands}
}

@inproceedings{hopcroft1971,
  author    = {Hopcroft, John E.},
  title     = {An n log n Algorithm for Minimizing States in a Finite Automaton},
  booktitle = {Theory of Machines and Computations},
  editor    = {Kohavi, Zvi and Paz, Azaria},
  publisher = {Academic Press},
  address   = {New York},
  year      = {1971},
  pages     = {189--196}
}
\end{document}